# Very small insects use novel wing flapping and drag principle to generate the weight-supporting vertical force

Xin Cheng and Mao Sun*

Institute of Fluid Mechanics, Beijing University of Aeronautics & Astronautics, Beijing 100191, China (*m.sun@buaa.edu.cn)

The effect of air viscosity on the flow around an insect wing increases as insect size decreases. For the smallest insects (wing length $R$ below 1 mm), the viscous effect is so large that lift-generation mechanisms used by their larger counterparts become ineffective. How the weight-supporting vertical force is generated is unknown. To elucidate the aerodynamic mechanisms responsible, we measure the wing kinematics of the tiny wasp *Encarsia formosa* (0.6 mm $R$) in hovering or very slow ascending flight and compute and analyze the aerodynamic forces. We find that the insects perform two unusual wing-motions. One is "rowing": the wings move fast downward and backward, like stroking oars; the other is the previously discovered Weis-Fogh 'fling'. The rowing produces 70% of the required vertical force and the Weis-Fogh 'fling' the other 30%. The oaring wing mainly produces an approximately up-pointing drag, resulting in the vertical force. Because each oaring produces a starting flow, the drag is unsteady in nature and much greater than that in steady motion at the same velocities and angles of attack. Furthermore, our computation shows that if the tiny wasps employed the usual wing kinematics of the larger insects (flapping back and forth in a horizontal plane), vertical force produced would be only 1/3 of that by the real wing kinematics; i.e. they must use the special wing movements to overcome the problem of large viscous effects encountered by the commonly used flapping kinematics. We for the first time observe very small insects using drag to support their weight and explain how a net vertical force is generated when the drag principle is applied.

## 1. Introduction

Although the wing of an insect beats at high frequency (usually above 100 Hz), the velocity of the wing relative the undisturbed air is small, owing to the small wing-length. As a result, the vertical force coefficient of the wing required to balance the weight is relatively high; the mean vertical force coefficient required is around 2 (Ellington 1984*a-d*; Sun & Du 2003), about three times as large as that of a cruising airplane. This high vertical-force coefficient cannot be explained by conventional aerodynamics (which applies to the steady-state, high-Reynolds number and attached flow).

Much work has been done to reveal the generation mechanisms of the high vertical-force coefficient (Sane 2003; Wang 2005; Shyy *et al.* 2010). It has been shown that insects use the lift of the wings to provide the weight-supporting vertical force (i.e. using lift principle for flight) and that the high lift is produced mainly by the leading-edge vortex (LEV) that attaches to the wing during an entire down- or upstroke, which is referred to as the delayed-stall mechanism (Ellington *et al.* 1996; Dickinson, Lehmann & Sane 1999; Liu et al. 1998; Bomphrey, Srygley & Thomas 2002). The LEV, together with the root vortex, the trailing-edge vortex and the tip vortex, form a vortex loop. As the wing moves, the attached LEV facilitates a linear growth of the vortex loop with time, producing a large time rate of change of the first moment of vorticity (or fluid impulse), hence a large aerodynamic force (Sun & Wu 2004). Because of the LEV's essential role in the high-lift generation of a flapping wing at low *Re*, in recent years much work has been conducted to study its fluid-dynamic properties. These studies have revealed the detailed structure of the LEV and how it changes with the Reynolds number (e. g. Birch, Dickson & Dickinson 2004; Kim & Gharib 2010; Ozen & Rochwell 2012; Jardin, Farcy & David 2012; Garmann, Visbal & Orkwis 2013), the aspect ratio of wing (e. g. Luo & Sun 2005; Carr, Chen & Ringuette 2013, Harbig, Sheridan & Thompson 2013) and the wing shape (Wolfinger & Rockwell 2014), and have partially explained the stability of the LEV [here the word 'stability' is used to describe whether the LEV remains attached throughout the stroke (stable) or if it is shed (unstable); e. g. Lentink & Dickinson 2009; Shyy *et al.* 2009; Wojcik & Buchholz 2014].

However, the above results do not apply to the smallest insects because of the very large effect of air viscosity. The viscous effect on the flow around an insect wing depends on the Reynolds number ($Re$), which is the ratio between inertial and viscous forces acting on a volume of air. The smaller the $Re$ is, the larger the viscous effect. $Re$ is proportional both to the chord length and the relative velocity of the wing. As insect size becomes smaller, both these two values decrease. $Re$ therefore decreases with the decreasing insect-size. For the smallest insects such as the tiny wasp *Encarsia Formosa* (wing length is about 0.6 mm), $Re$ is about 10 (Weis-Fogh, 1973). At this low $Re$, the viscous effect is so large that the LEV is significantly defused and little lift can be generated, while the drag is large (Miller & Peskin 2004; Wu & Sun 2004). Therefore, the smallest insects cannot use the same lift generation mechanisms as that of their larger counterparts, and they may even do not use the lift principle for flight. That is, the smallest insects may use a drag mechanism to for flight. That very small insects may use drag rather than lift for flight was suggested a long time ago by Horridge (1956). But there were some arguments against it (Weis-Fogh, 1975). First, the drag principle was difficult to apply for the following reason: at very low $Re$ the drag of a wing tended to become independent of the angle of attack, hence the drag of the effective stroke might be cancelled by that of the return stroke, producing little net force. Second, people had never yet observed a small insect in flapping flight that used drag rather than lift. Weis-Fogh (1973) found that the tiny wasp *Encarsia formosa* performed an unusual wing motion, 'clap and fling', and it was shown that the 'fling' can produce a momentary large vertical force (Spedding & Maxworthy 1986). But because detailed quantitative data on wing kinematics was not readily available at the time, it was not known if the vertical force generated by the Weis-Fogh 'fling' was enough for weight support.

So far, how the smallest insects produce the weight-supporting vertical force, whether or not they use the drag principle for flight, and if drag principle is used, how a net vertical force is generated at the extremely low $Re$, remain unknown.

In the present study, we first used high-speed cameras equipped with micro-lenses and extension tubes to measure the detailed wing kinematics in the tiny wasp *Encarsia formosa*, and then used a well-tested flow solver to solve the Navier-Stokes equations and obtained the flows and aerodynamic forces. We have achieved the following results: We find that in addition to the Weis-Fogh 'fling', there is another special wing kinematics, "rowing": the wings stroke fast downward and backward at a large angle of attack, like the movement of the oars of a boat. It is shown that the rowing produces 70% of the required vertical force, and the Weis-Fogh 'fling' produces the other 30%. It is shown that the large vertical force by the rowing is generated by an unsteady-drag mechanism. We for the first time observe tiny insects using drag to support their weight and explain how a net vertical force is generated when drag principle is applied.

## 2. Experimental and computational methods

### 2.1 *Experimental systems and techniques*

*Encarsia formosa* Gahan were acquired from the Laboratory of Biological Invasion of Institute of Plant Protection, Chinese Academy of Agricultural Sciences, which were descendents of wild-caught *E. formosa*. The wasps were housed in small chambers in groups of three to five individuals and fed with 20% sugar solution. The flight of the wasps in a transparent flight chamber (40×40×30 mm$^3$) was filmed using three orthogonally aligned synchronized high-speed cameras (FASTCAM Mini UX100, Photron Inc., San Diego, CA, USA; 10,000 frames per second, shutter speed 20μs, resolution 896×488 pixels) mounted on an optical table (Figure 1). Each camera was equipped with a 60 mm micro-Nikkor lens and 12 mm extension tube. Each camera view was backlit using a 50 W integrated red light emitting diode (LED; luminous flux, 4000 lm; wavelength, 632 nm). The light was made uniform by two lenses. We manually triggered the synchronized cameras when the insect was observed to fly steadily in the filming area of approximately 4×4×4 mm$^3$, which represented the intersecting field of views of the three cameras. The experiment was conducted at room temperature 25-27°C and relative humidity 50-60%.

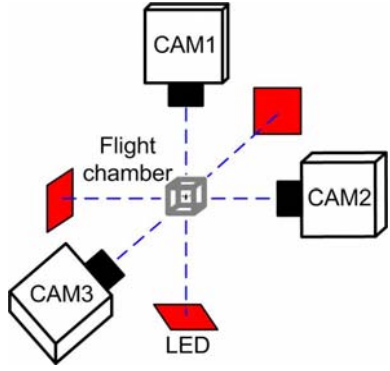

FIGURE 1. A sketch showing the experimental setup: the flight chamber and cameras; each camera view is backlit using a integrated red light emitting diode (LED).

The method used to extract the 3D body and wing kinematics from the filmed data was a modification of that used in several previous works of our group (Liu & Sun 2008; Mou, Liu & Sun 2011). The original method and the modification are described in Appendix A.

### 2.2 *Methods of flow and aerodynamic-force computation*

The flows around and the aerodynamic force acting on the insect were computed using the method of computational fluid dynamics. The governing equations of the flow are the incompressible Navier-Stokes equations, the dimensionless form of which is:

$$\nabla \cdot \mathbf{u} = 0, \qquad (2.1)$$

$$\frac{\partial \mathbf{u}}{\partial \tau} + \mathbf{u} \cdot \nabla \mathbf{u} = -\nabla p + \left(\frac{1}{Re}\right)\nabla^2 \mathbf{u}. \qquad (2.2)$$

where $\mathbf{u}$ is the non-dimensional fluid velocity, $p$ the non-dimensional fluid pressure, $\tau$ the non-dimensional time, $\nabla$ the gradient operator, $\nabla^2$ the Laplacian operator and $Re$ the Reynolds number; in the non-dimensionalization, $c$, $U$ and $c/U$ are taken as reference length, velocity and time, respectively (here $c$ is the mean chord length of wing and $U$ is the mean flapping velocity at the radius of gyration of wing).

In the present study, the linked fore- and hindwings were modeled as a single wing; but the forewing could rotate relative to the hindwing around the "leading-edge" of the hindwing, which happened during the 'fling'. The wing of *E. formosa* consists of the membrane and the brim of marginal hairs (Figure 2a), but here the hairs that line the wing are modeled as an extension of the membranous wing, i.e. the planform of the model wing is similar to that of a wing consisted of the membrane and the brim of marginal hairs (Figure 2b). The effects and limitations of these simplifying assumptions will be discussed in a later section (§7). The section thickness of model wing is 3% of the local chord length and the leading and trailing edges are rounded. The body shape is given by fitting ellipses to the body in the video images.

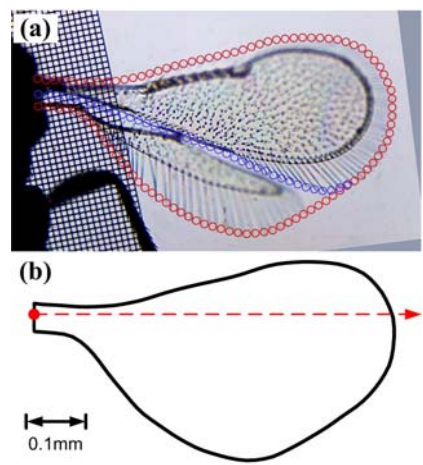

FIGURE 2. (a) Wing planform of an *Encarsia formosa*, extracted from microscope images. Blue circles indicate the boundary between fore- and hindwings. (b) The planform of the model wing used in computations. The red dot indicates the wing base and the dashed line indicates the axis of pitch rotation.

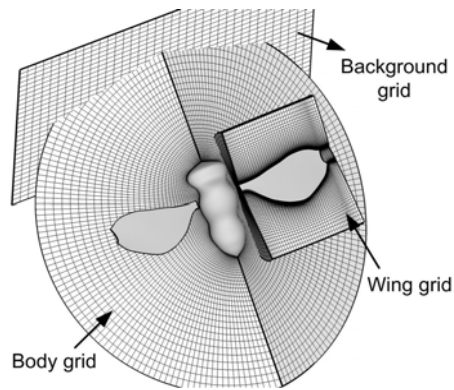

FIGURE 3. Portions of computational grids.

The equations were solved over moving overset grids because there are relative movements between the left and right wings and between the body and wing. There was a body-fitted curvilinear grid for each of the wings and the body and a background Cartesian grid which extends to the far-field boundary of the domain (Figure 3). In the calculation process, the wing grid was regenerated in every time-step because of the spanwise bending deformation and the rotation of the forewing relative to the hindwing during the 'fling'. The computational method is the same as that used in several previous studies of our group (Sun & Yu 2003, 2006; Du & Sun 2012) and only an outline of the method is given here. The algorithm is based on the method of artificial compressibility. It was first developed by Rogers, Kwak & Kiris (1991) for single-grid, and then extended by Rogers & Pulliam (1994) to overset grids. The time derivatives of the momentum equations are differenced using a second-order, three-point backward difference formula. To solve the time discretized momentum equations for a divergence free velocity at a new time level, a pseudo-time level is introduced into the equations and a pseudo-time derivative of pressure divided by an artificial compressibility constant is introduced into the continuity equation. The resulting system of equations are iterated in pseudo-time until the pseudo-time derivative of pressure approaches zero, thus, the divergence of the velocity at the new time level approaches zero. The derivatives of the viscous fluxes in the momentum equation are approximated using second-order central differences. For the derivatives of convective fluxes, upwind differencing based on the flux-difference splitting technique is used. A third-order upwind differencing is used at the interior points and a second-order upwind differencing is used at points next to boundaries. With overset grids, the solution method for single-grid is applied to each of the wing grid and the background grid, and data are interpolated from one grid to another at the inter-grid boundary points using tri-linear interpolation. Details of this algorithm can be found in Rogers *et al.* (1991) and Rogers & Pulliam (1994). For the far-field boundary conditions, at the inflow boundary, the velocity components are specified according to relative velocity at the boundary while pressure is extrapolated from the interior; at the outflow boundary, pressure is set equal to the static pressure of the still air and the velocity is extrapolated from the interior. On the wing surfaces, impermeable wall and no-slip boundary conditions are applied, and the pressure on the boundary is obtained through the normal component of the momentum equation written in the moving coordinate system. The background Cartesian grid was generated algebraically. The wing grid (O-H type) and body grid (O-O type) were generated using a Poisson solver which was based on the work of Hilgenstock (1988). The grids will be further described in §3.1, as will be the analysis of the convergence of solutions.

## 3. Measured wing kinematics

Flights in eight wasps were filmed; they are denoted as EF1, EF2, EF3, EF4, EF5, EF6, EF7 and EF8, respectively. Figure 4 shows the sequences of the flight of one of the flies (EF5); the original video sequences for the insect is presented as supplementary material, Movies 1-3.

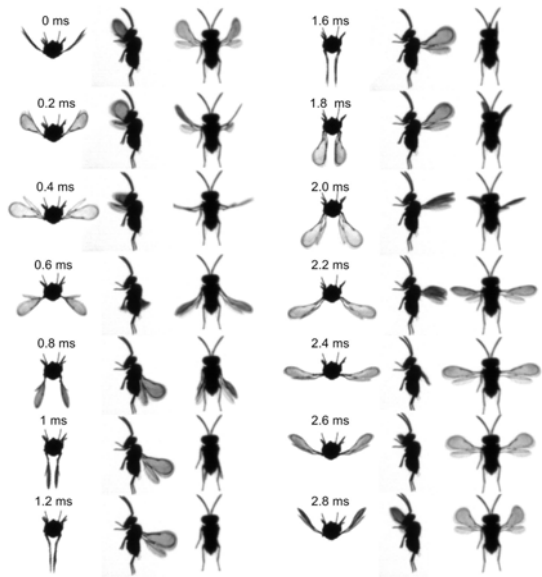

FIGURE 4. Video sequences of a *Encarsia formosa* at near-hovering; three camera views are shown. Times noted are ms from the instant when the rowing starts. For complete video sequences, see Movies 1-3.

In describing the wing kinematics of the flies, a method similar to that given by Ellington (1984*b*) was used. For each insect, data of six wingbeat cycles were processed. In each cycle, when the wings were at the foremost and rearmost positions, the wing-tip points were projected onto the plane of symmetry of the insect. A linear regression line of the projections was then determined. The plane which is parallel to the above line and passes the wing base is defined as stroke plane (Figure 5). Let (*X*, *Y*, *Z*) be a reference frame with origin at the wing base, the *X* axis being horizontal and pointing backward, the *Z* axis being vertical and pointing upward and the *Y* axis pointing sideward (Figure 5). The orientation of a wing relative to the stroke plane can be determined by three Euler angles (Figure 5): $\phi$, positional angle of wing (in the stroke plane); $\psi$, pitch angle; $\theta$, deviation angle. $\beta$, stroke-plane angle.

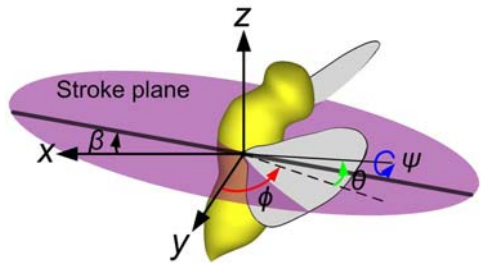

FIGURE 5. Reference frame and Euler angles defining the wing kinematics.

Figure 6 gives the measured Euler angles, $\phi$, $\psi$ and $\theta$, and the spanwise bending in one wingbeat cycle in a flying *E. formosa* (EF1). Based on the data given in Figure 6, the stroke diagrams of the wing motion are plotted and shown in Figure 7. Larger insects in normal hover-flight sweep their wings back-and-forth horizontally (inset, Figure 7a; Ellington 1984b). For the tiny wasp, the mid-portion of the forth sweep is similar to that in normal hovering; however, towards the end of the forth sweep, the wings are swung upwards to a higher position. Then the wings move fast downward and backward with the wing-surface approximately horizontal, the motion resembles that of the stroking oars of a boat or the fins of a fish swimming upwards. We thus see that the "back sweep" of the wasp is greatly different from that of larger insects in normal hovering. Near the end of the "back sweep", the wings 'clap' and move vertically upwards until the wing-tips reaching the level of the head (Figure 7a, b), then the wings 'fling' open. This is the 'clap' and 'fling' described previously by Weis-Fogh (1973). Wing motions of another seven individuals (Table 1; Movie 1 in supplemental material) were measured. Most of the insects filmed were in slow ascending flight (e.g. EF1, Movie 1) and their advance ratio $J$ (the velocity of body divided by the mean wing-tip speed) ranged from 0.02 to 0.12 (Table 1). The one with $J$=0.02 (EF5, Movie 3) can be considered as in 'true' hovering. Note that all the insects have approximately the same wing kinematic pattern (see Figures A2-A4 in Appendix B).

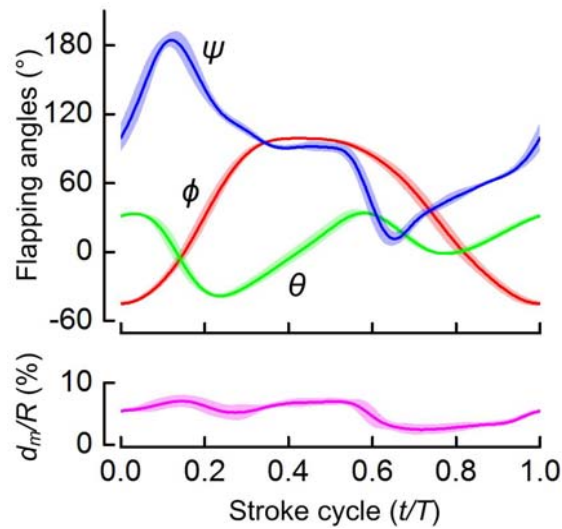

FIGURE 6. Measured Euler angles and the maximum bending displacement ($d_m/R$) from one individual, EF1 (mean±s.d.; $n$=6 wingbeats). $T$, stroke period; $R$, wing length.

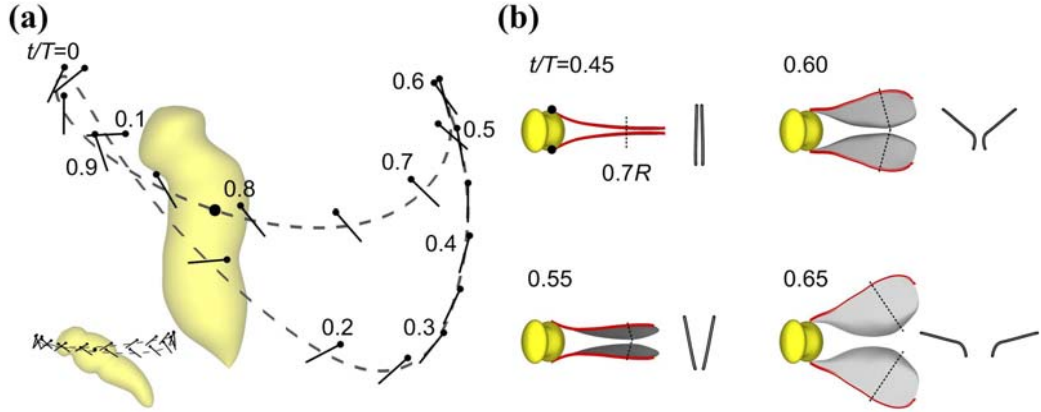

FIGURE 7. (a) Stroke diagrams show the wing motions of *Encarsia formosa* and dronefly (inset). Dashed curve indicates the wing-tip trajectory (projected onto the *X-Z* plane); black lines indicate the orientation of the wing at 20 temporally equidistant points, with dots marking the leading edge; black dot defines the wing-root location on the insect body. (b) Wing motion at dorsal stroke-reversal, showing the Weis-Fogh 'clap-and-fling'. *R*, wing length.

Other morphological data of the wasps required for carrying out the analysis were also measured in the experiment, or computed on the basis of the measured data. The data included (Table 1): the wing length ($R$), the area of one wing ($S$) and the distance between the two wing roots ($l_r$); the radius of the second moment of wing area or radius of gyration ($r_2$) is $0.64R$ for the eight insects. The weight of the insect could not be measured due to the limitation of the accuracy and measuring range of our balance.

Table 1. Flight parameters for the eight *Encarsia formosa*.

| ID | $J$ | $n$ (Hz) | $\Phi$ (°) | $R$ (mm) | $S$ (mm²) | $l_r$ (mm) | $Re$ | $F_w$ (μN) | $F_{w,h}$ (μN) | $\eta$ (°) | $D_b$ (μN) | $P_m$ (μW) |
|---|---|---|---|---|---|---|---|---|---|---|---|---|
| EF1 | 0.10 | 361±7 | 144±2 | 0.61 | 0.14 | 0.21 | 10.6 | 0.196 | 0.01 | 2.9 | 0.01 | 0.476 |
| EF2 | 0.08 | 362±7 | 149±2 | 0.59 | 0.13 | 0.22 | 10.2 | 0.172 | 0.016 | 5.4 | 0.007 | 0.431 |
| EF3 | 0.11 | 365±7 | 146±1 | 0.59 | 0.13 | 0.23 | 10.1 | 0.200 | 0.013 | 3.8 | 0.011 | 0.505 |

| | | | | | | | | | | | |
|---|---|---|---|---|---|---|---|---|---|---|---|
| EF4 | 0.10 | 360±6 | 148±1 | 0.61 | 0.14 | 0.22 | 10.8 | 0.210 | 0.002 | 0.5 | 0.01 | 0.513 |
| EF5 | 0.02 | 349±5 | 145±2 | 0.62 | 0.14 | 0.21 | 10.6 | 0.246 | 0.038 | 9.4 | 0.002 | 0.552 |
| EF6 | 0.05 | 345±0 | 142±2 | 0.62 | 0.14 | 0.21 | 10.5 | 0.230 | 0.043 | 10.8 | 0.006 | 0.498 |
| EF7 | 0.12 | 336±6 | 141±2 | 0.59 | 0.13 | 0.19 | 9 | 0.205 | 0.034 | 9.5 | 0.011 | 0.420 |
| EF8 | 0.11 | 345±0 | 149±1 | 0.57 | 0.12 | 0.20 | 9.3 | 0.169 | 0.013 | 4.5 | 0.009 | 0.400 |

*J*, advance ratio, defined as the velocity of body motion divided by the mean wing-tip speed; small *J* indicates that the insects are at near-hovering. *n* and Φ, stroke frequency and amplitude, respectively (mean ± s.d.). *R*, wing length; *S*, area of wing; $l_r$, the distance between the left and right wing-roots; *Re*, Reynolds number; $F_w$ and η, magnitude and angle from the vertical of the computed cycle-mean force vector of the wings, respectively; $F_{w,h}$, the cycle-mean horizontal force of the wings; $D_b$, body drag; $P_m$, cycle-mean power.

## 4. Flows and aerodynamic-force mechanisms

To assess how the weight-supporting force is generated by the special wing motions, in this section, the flow and forces on the wings were computed and analyzed.

### 4.1 *Code validation and grid test*

The flow solver had been tested by unsteady aerodynamic forces measured from translating (Meng & Sun 2013) and rotating (Sun & Yu 2006) model insect-wings. These tests showed that the unsteady aerodynamic forces computed by the present solver agreed well with the experimental measurements. But these tests were for cases of higher Reynolds number (*Re* above 200). Here we further test at lower Reynolds numbers (*Re*~10) identical to the wasp flight. First, we made a conceptual test for a single, flapping wing at *Re*=10 (the flapping motion is an idealized flapping motion often used in insect flight study, given by Dickinson et al. 1999). We computed the flow first using the single-grid, and then computed the flow using the multi-block, overset-grids. Since the physical problem is the same in the two computations, the results should be the same. This can be a test on the multi-block, overset solver. Figure 8a shows the comparison in the lift and drag coefficients ($C_L$ and $C_D$) between the two computations. There is very good agreement between these two sets of results, showing that the multi-block, overset Navier-Stokes solver works well at *Re* around 10.

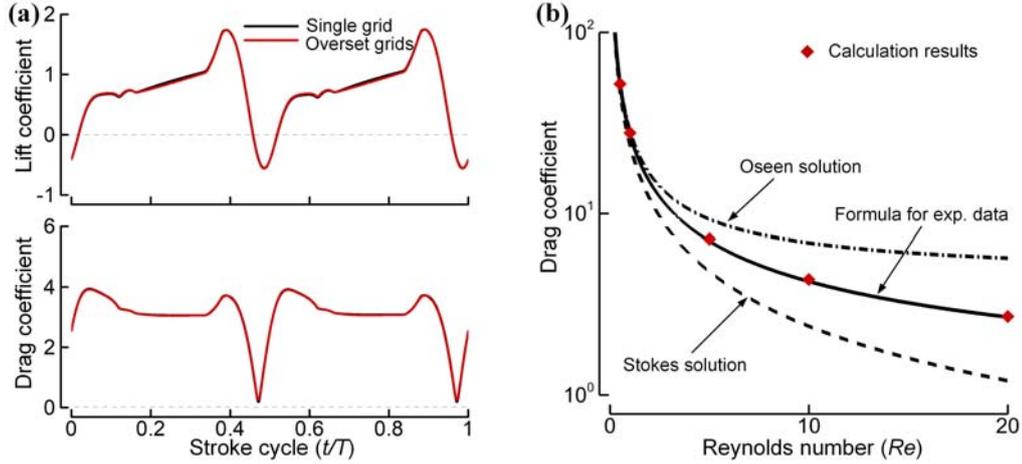

FIGURE 8. (a) Lift and drag coefficients of a single wing flapping at *Re*=10 computed using a single-grid, compared with those computed using the multi-block, overset-grids. (b) The calculated drag coefficient of a sphere at *Re*=0.5-20, compared with the theoretical results and experimental data.

To further validate the low-*Re* simulation, we computed the flows around a sphere at *Re*=0.5-20, for which there exist Stokes or Oseen solutions and accurate experiment data (White 1991). Figure 8b compares the computed drag coefficient with the theoretical and experimental data. The difference between the computed results and the data is less than 2%.

As mentioned above the flow solver was tested by measured data of translating and rotating wings. Recently, lift coefficient ($C_L$) was measured for model insect-wings in flapping motion at *Re*=180 (Han, Chang, & Han 2016) and in revolving motion with an acceleration phase, a constant speed phase and a deceleration phase *Re*=500 (Manar & Jones 2014). Here we make a further test of the solver using these experiment data. As seen in Figure 9, the computed $C_L$ agrees very well with the measured.

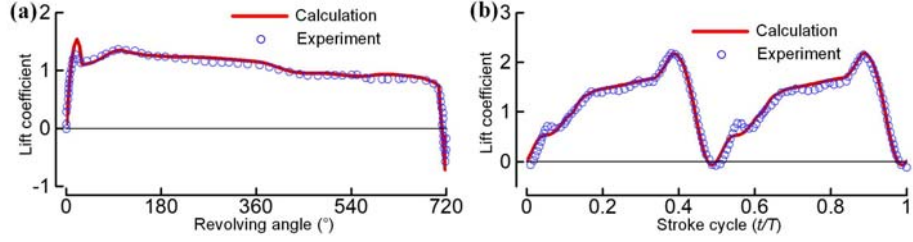

FIGURE 9. Comparison between the calculated and measured lift coefficients on a rectangular wing in revolving motion with an acceleration phase, a constant speed phase and a deceleration phase ($Re$=500) (a), and on a model fruitfly wing in flapping motion ($Re$=180) (b).

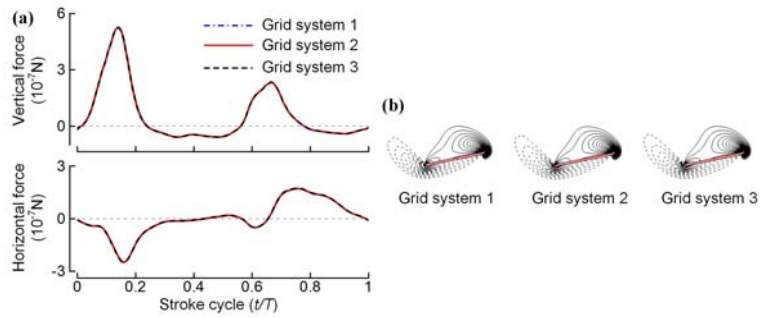

FIGURE 10. The moving overset grids and grid resolution test. (a) Time courses of the vertical and horizontal forces of the wings in one cycle, calculated with three grid-systems. (b) The corresponding non-dimensional spanwise vorticity contours at section $0.7R$ at $t/T$=0.14; the magnitude of the non-dimensional vorticity at the outercontour is 3 and the contour interval is 1.

Before proceeding to compute the flows, grid resolution tests were conducted to ensure that the flow calculation was grid independent. Three grid-systems were considered. For grid-system 1, the wing grid had dimensions $41\times 61\times 43$ in the normal direction, around the wing, and in the spanwise direction, respectively (first layer grid thickness was $0.0015c$); the body grid had dimensions $66\times 51\times 35$ along the body, in the azimuthal direction and in the normal direction, respectively; the background grid had dimensions $81\times 81\times 81$ in the $X$, Y and $Z$ directions, respectively. For grid-system 2, the corresponding grid dimensions were $61\times 91\times 65$, $99\times 77\times 53$ and $121\times 121\times 121$ ($0.001c$). For grid-system 3, the corresponding grid dimensions were $91\times 135\times 96$, $149\times 118\times 80$ and $181\times 181\times 181$ ($0.00067c$). For all the three grid-systems, the outer boundary of the background grid is $20c$ from the wing root; the grid points of the background grid concentrated in the near field of the wings where its grid density was approximately the same as that of the outer part of the body-grid. The typical cell size in the near wake where vortices from a stroke may interact with the wing in the subsequent strokes is less than $0.038c$, $0.061c$ and $0.1c$ for grid-systems 3, 2 and 1, respectively. The aerodynamic forces and vorticity patterns computed from three grid systems are shown in Figure 10 (the wings started to flap in a quiescent fluid, and after about three cycles the flow became periodical). It is observed that there is almost no difference between the forces calculated by the three grid-systems (Figure 10a). The first grid refinement produces a little change in the vorticity plot, and the second grid refinement produces almost no change (Figure 10b). Calculations were also conducted using a larger computational domain. The domain was enlarged by adding more grid points to the outside of the background grid of grid-system 2 (the outer boundary of the background grid of grid-system 2 was increased from $20c$ to $40c$). The calculated results showed that there was no need to put the outer boundary further than that of grid-system 2. The non-dimensional time step was 0.02 (non-dimensionalized by $c/U$). The effect of time step value was studied and it was found that a numerical solution effectively independent of the time step was achieved if the time step value was ≤0.02. From the above discussion, it was concluded that grid-system 2 and time step was 0.02 were proper for the calculation.

### 4.2 *Aerodynamic forces*

In the present study, the vertical and horizontal components of the total aerodynamic force of a wing are referred to as vertical force and horizontal force, respectively. Because the wing does not translate in a horizontal plane, the vertical force is not the lift of the wing and the horizontal force not the drag. The velocity at the radius of gyration ($r_2$) of the wing is used to represent the velocity of the wing (Figure 11e). Lift and drag are defined as the components of the total aerodynamic force that are perpendicular and parallel to the velocity of the wing, respectively. Figure 11a-d give the time histories of the computed aerodynamic forces of a wing in one cycle for EF1 (results for the other seven individuals are given in Appendix B, see Figure A5). Because the computed forces show great consistency for the eight individuals, we analyze the results of EF1 (Figure 11) as an example. As seen from the figure, a large vertical force peak is produced during the rowing (Figure 11a, $t/T≈0$-$0.2$) and another smaller one during the 'fling' (Figure 11a, $t/T≈0.55$-$0.7$); the rest parts of the cycle do not produce any positive vertical force. From the data in Figure 11a, it is calculated that the rowing contributes 70% of the total vertical force and the 'fling' the rest 30%. The rowing produces a forward pointing horizontal force and the 'fling' and the subsequent forth sweeping produce a backward pointing horizontal force (Figure 11b), and the two approximately cancel, as required for hovering and ascending flight.

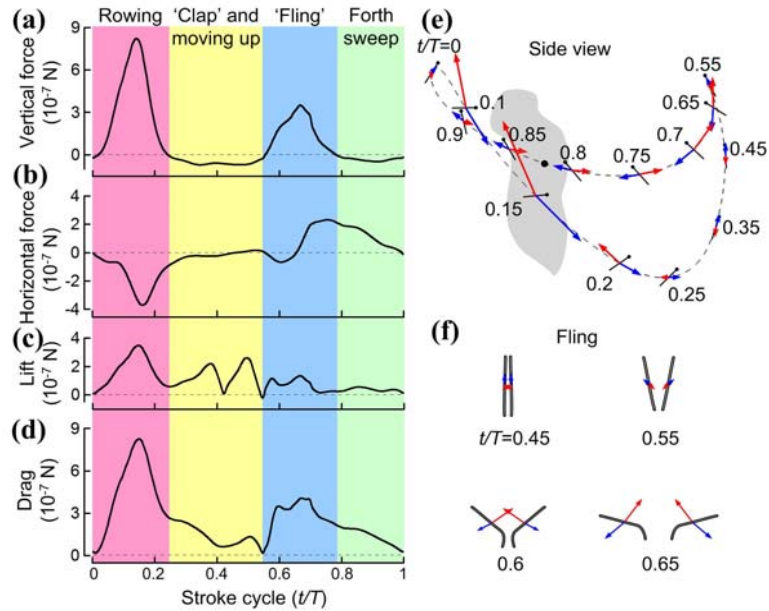

FIGURE 11. Aerodynamic forces. (a) to (d), vertical force, horizontal force, lift and drag of a wing. (e) Diagram showing the wing position, velocity vector and force vector at various times in one stroke cycle (side view). (f) Back view of the wing in 'fling' motion, represented by wing section at $0.7R$, and the corresponding velocity vector and force vector. Blue arrow, velocity; red arrow, force.

As aforementioned, the insects are in hovering or very slow ascending flight. Thus the computed mean force vector of the wings should be vertical. As seen in Table 1, the mean force vector is very close to being vertical: its horizontal component ($F_{w,h}$) is small and the angle between the force vector and the vertical (η) is less than 6º for five of the eight wasps and about 10º for the other three. The magnitude of the mean force vector of the wings ($F_w$) is shown in Table 1. When in slow ascending flight, there is a small drag on the body (the computed body drag, $D_b$, is also given in Table 1). The value of ($F_w$-$D_b$) should equal to the weight of the insect. As mentioned above, we could not measure the weight of the insects in our experiment. Weis-Fogh (1973) measured the weight of 10 wasps of the same species as those in our experiment (*E. Formosa*) and gave the average weight of the insects. Since the wasps in our experiment and those in Weis-Fogh's experiment are the same species, they are geometrically similar. Thus we can make an indirect comparison here. For the insects in Weis-Fogh's experiment, the average wing length was 0.62mm, the average wing-tip speed was 1.17m/s and the average weight was 0.245μN. For the eight insects in our experiment, the average wing length is 0.6mm and average wing-tip speed is 0.1075m/s. Assuming that the aerodynamic force is proportional to the wing area (or the square of the wing length) and to the square of wing-tip speed, the average weight of the insects in our experiment can be estimated as

$$(\frac{0.6}{0.62})^2 \times (\frac{1.075}{1.17})^2 \times 0.245\mu N = 0.194\mu N \qquad (4.1)$$

The computed average $F_w$-$D_b$ for the eight insects is 0.196μN, in good agreement with the above estimated average weight (0.194μN).

### 4.3 *Aerodynamic-force mechanisms*

The vertical and horizontal forces of a wing come from its lift and drag. The rowing produces a large drag and a much smaller lift (Figure 11c, d). Therefore, the large vertical force by the rowing is mainly from the drag of the wing (approximately 70% of the vertical force during the rowing period is contributed by the drag). In this period, the wing is in a transient motion (velocity increases from zero at *t/T*=0 to a large value at *t/T*=0.15 and then decreases to a small value at *t/T*≈0.2) and the angle of attack is rather large, between 46° to 65° (Figure 11e). The transient motion and force here are very similar to those of an oar of a boat. Let us examine how the large drag is produced.

Since the wing is in fast acceleration ($t/T$=0-$t/T$≈0.15) and deceleration ($t/T$≈0.15-$t/T$≈0.2), we first look at the role of added mass. Howe (1989, 1995) made an analysis of the force exerted on a rigid body in unsteady motion in a uniform incompressible viscous fluid, and showed that the force on the body is the summation of three parts. The first part is caused by the acceleration of the body and the expression of this part is identical to the added mass force obtained by solving the potential flow of no free vorticity, and the second and third parts are due to the free vorticity in the wake and the viscous skin friction, respectively. His results show that the added mass of a rigid body moving in an incompressible fluid can be computed by solving a separate potential flow. Our group recently developed a code for computing the added mass force of flapping wings (Liu and Sun 2018). Using the code we calculated the added mass force during the rowing period and the results are shown in Figure 12a and b. It is seen that the added mass forces are relatively small. The large drag must be produced by other mechanisms.

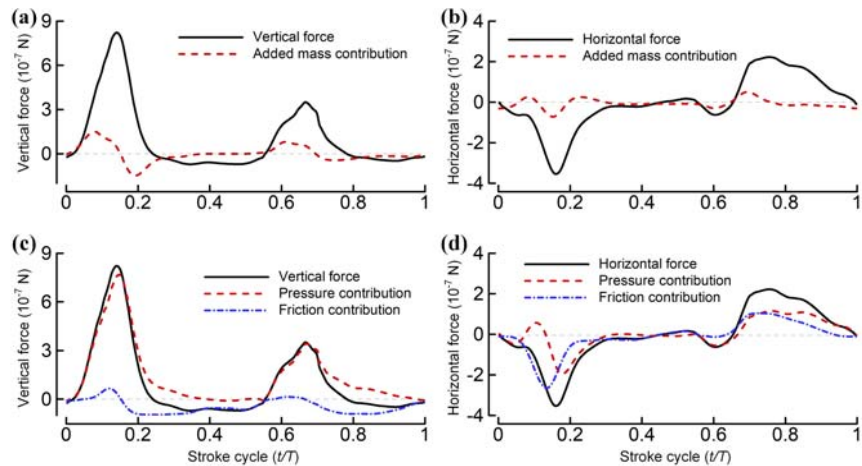

Figure 12. (a) and (b) Added-mass contributions to the vertical and horizontal forces. (c) and (d) The contributions of the pressure- and friction-based components to the vertical and horizontal forces.

Figure 12c and d shows the contributions of the pressure- and friction-based components to the vertical and horizontal forces. It is seen that the large vertical force or the large drag of the wing is mainly contributed by the pressure-based component (Figure 12c). Let us investigate how the large pressure-based force is generated. Flow field data in this period are shown in Figure 13. The fluid is pushed downwards by the "lower" wing surface and sucked downwards by the "upper" wing surface (Figure 13c). Strong vortices are shed at the leading and trailing edges (Figure 13a, b). The pushing results in a large pressure on the lower surface of the wing and the induced velocity of the leading-edge vortex (LEV) and trailing-edge vortex (TEV) result in a large suction pressure on the upper surface of the wing (Figure 13d), giving the large drag or vertical force. Note that for the tiny insects here, both the LEV and TEV are used in enhancing the aerodynamic production, unlike the cases for larger insects whose *Re* is larger than 100, where LEV dominates the enhancement of aerodynamic production. We thus see that the large pressure force that gives the large drag or vertical force is generated by the unsteady effect similar to that of starting flow of a plate started in its normal direction.

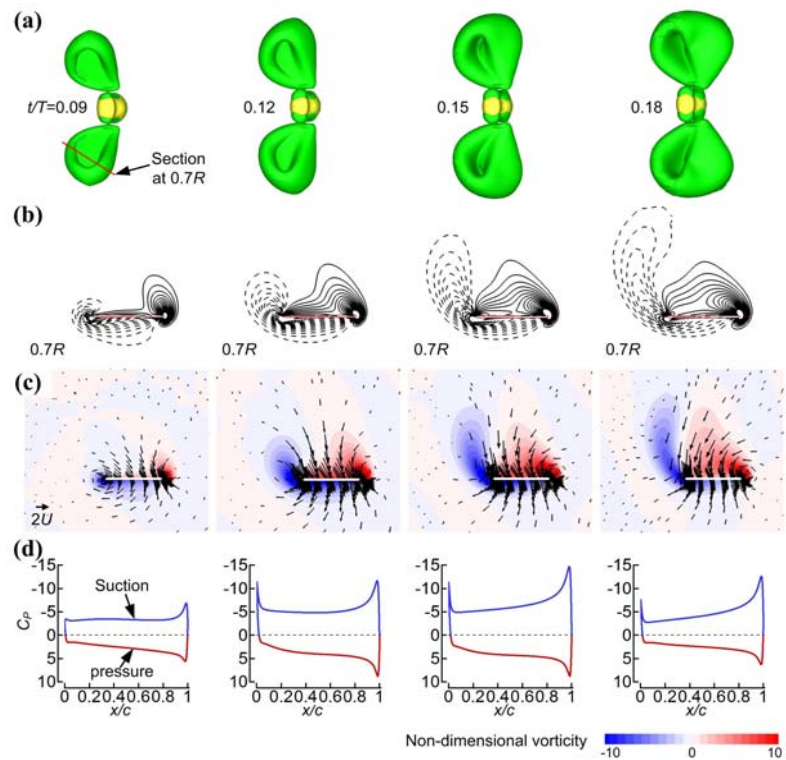

FIGURE 13. Flow fields and surface pressure distributions during the rowing period. (a) Iso-vorticity surface plots (top view) at $t/T$=0.09, 0.12, 0.15 and 0.18 (in this and other plots of iso-vorticity surface, the magnitude of the non-dimensional vorticity is 2). Vorticity is nondimensionalized by $U/c$. (b) Non-dimensional spanwise vorticity contours at section $0.7R$. solid and dashed lines denote the anticlockwise and clockwise vorticity, respectively. The magnitude of the non-dimensional vorticity at the outercontour is 1 and the contour interval is 1. (c) Non-dimensional velocity vector plot at section $0.7R$. (d) Nondimensional surface pressure distributions at section $0.7R$. Non-dimensional pressure, $C_P$, is defined as $(p-p_\infty)/0.5\rho U^2$.

To further show the unsteady effect of the transient motion in the rowing period, we computed the corresponding steady flows of the wing, i.e. at each time point in the rowing period, we use the instantaneous speed and angle of attack of the wing at that time to compute a steady-state flow. For example, at $t/T$=0.09 in Figure 13, the wing's tip-speed and angle of attack were known; we let the wing move (revolve) at the given tip-speed and angle of attack until the flow reached steady state (after the wing revolved for more than one revolution, the flow became steady), thus we obtain the corresponding steady flow and force for this time point. We do the same for other time points, $t/T$ =0.12, 0.15 and 0.18 and so on. The steady-state results are shown in Figure 14. It is seen that in the steady-state cases, the vortices shed at the leading and trailing edges are much weaker and the pressure on the lower surface and the suction on the upper surface are much smaller than the corresponding ones in the transient motion (comparing results in Figure 14 with those in Figure 13). The reason for the LEV and TEV in the unsteady case are much stronger than those in the steady-state case is as following. The wing in the unsteady case is similar to a plate in starting motion in its normal direction, in which vorticity shed at the edges of the plate could accumulate to a large amount behind the plate. We computed the vertical force from the steady-state case and it is 45% less than that produced by the unsteady (rowing) motion.

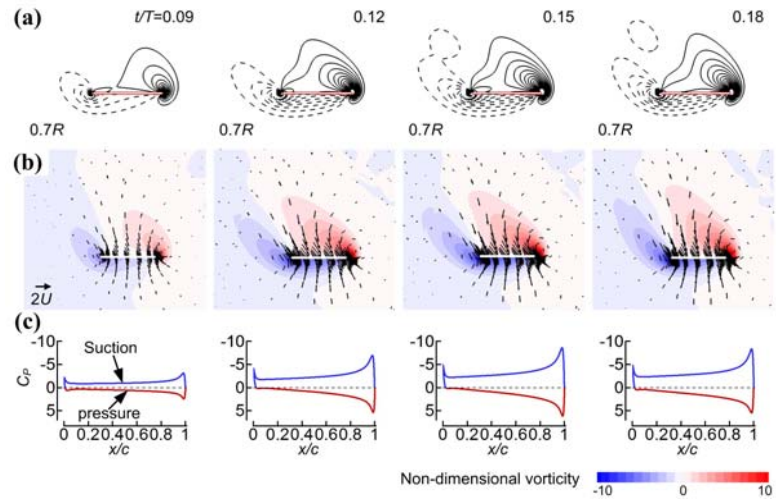

FIGURE 14. Flow fields and surface pressure distributions for the quasi-steady case, corresponding to those in Figure 13. (Data here are non-dimensionalized using the same velocity as that used in Figure 13.)

Next, let us look at the vertical force peak during the 'fling'. It is also due to the large drag on the wing (see Figure 11a, d and f). The production mechanism of the large drag has been explained in detail by many previous studies (Spedding & Maxworthy 1986; Sun & Yu 2003; Miller & Peskin 2005; Arora et al. 2014). Flow field data during the 'fling' are shown in Figure 13. As pointed out in previous studies (Spedding & Maxworthy 1986; Sun & Yu 2003; Miller & Peskin 2005), the mechanism of the vertical force peak during the 'fling' is that the opening of the wing-pair generates a vortex ring containing a strong downward-jet (see Figure 15a and c) in a very short period, or a large fluid impulse, resulting in the large force.

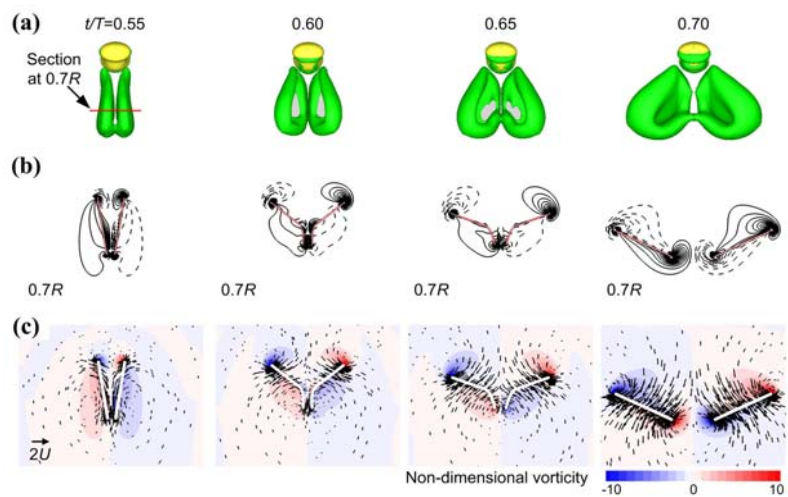

FIGURE 15. Flow fields during the 'fling' period. (a) Iso-vorticity surface plots (top view) at *t/T*=0.55, 0.60, 0.65 and 0.70. (b) Non-dimensional spanwise vorticity contours at section 0.7*R*. solid and dashed lines denote the anticlockwise and clockwise vorticity, respectively. (c) Non-dimensional velocity vector plot at section 0.7*R*.

## 5. Aerodynamic power

From the computed aerodynamic force and the wing motion data, the power used by the wing to generate the forces was calculated and shown in Figure 16. It is seen that power in the rowing and the 'fling' periods is large, as is expected.

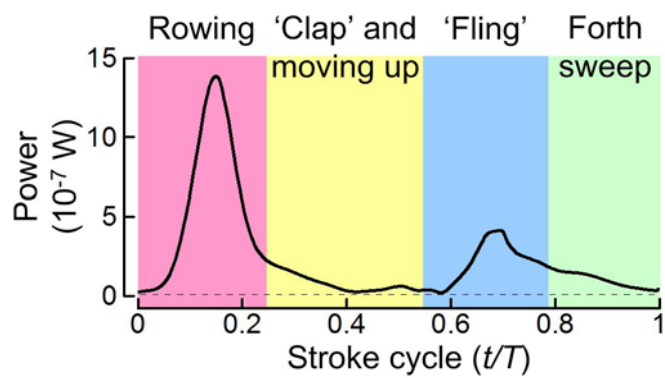

FIGURE 16. Aerodynamic power of a wing in one wingbeat cycle.

From data in Fig. 11a and Fig. 16, the mean vertical force and mean power are calculated as 0.196 μN and 0.476 μW respectively, giving a value of 2.43 W per Newton lifted (referred to as specific power). Weis-Fogh (1975) found that large (10 kg) or small (1 μg) animals derive mechanical power from the same type of muscle and the specific power that an animal can supply is independent of size, and it can be up to 6 W/N. The tiny wasp's required power (2.4 W/N) is far below this limit. Power is consumed by a wing when it moves against the drag. Airplanes, birds and larger insects use lift to support the weight and the specific power is made small by having a large lift-to-drag ratio (e.g. 10). For the tiny wasp, drag is used to support the weight and the "lift-to-drag" ratio is 1, far from being large. Why the tiny wasps still have a small specific power? Let *D* be the mean drag of one wing and it is equal to half of the weight.

The mean angular speed of the wing is proportional to the wingbeat frequency *n* and the torque due to the drag to *RD* (*R*, wing length). Then the power is proportional to *nRD* and the specific power to *nR*. Since *n* varies with size according to $R^{-0.68}$ (Dudley 2002), the specific power is proportional to $R^{0.32}$. As insect size decreases, the specific power would decrease, explaining the above question.

## 6. Why special wing movements and drag principle must be used

Using drag to produce the weight-supporting vertical force has been shown previously in some insects. Wang (2004) and Sun and Lan (2004) found that a hovering dragonfly, with inclined stroke plane and asymmetric flapping, used drag to support about three quarters of its weight. For a hovering hoverfly with inclined stroke plane and asymmetric flapping, about 45% of the vertical force was contributed by the drag (Zhu and Sun 2017). These insects (dragonfly and hoverfly) can also hover with a horizontal stroke plane (Ellington 1984b; Wakeling and Ellington 1997); in this case the vertical force is produced using the lift principle only. That is, dragonflies and hoverflies can choose to use lift or drag to produce the weight-supporting force. However, the tiny wasps must use the unusual wing motions and the drag principle to fly. To understand the reasons for this, we simplified its wing kinematics to that employed by the larger insects: wings sweeping back and forth in a horizontal plane (Figure 7a, inset), but kept *Re* unchanged and re-computed the flows and forces. In the computation, the stroke amplitude (144°), frequency (361 Hz), Reynolds number (10.6) are the same as those of the tiny wasp. The angle of attack at mid-stroke ($\alpha_m$) is set as 35° to 55° (as will be seen below, when $\alpha_m$ is larger than 40°, the lift will change little and drag become larger and larger). The computed results are shown in Figure 17.

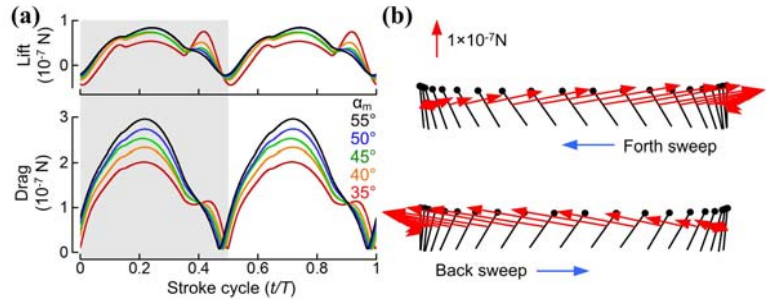

FIGURE 17. Aerodynamic forces produced if the tiny wasp used wing kinematics as that of the large insects. (a) Lift (vertical force) and drag. Grey shading represents the forth stroke. (b) Diagram of wing motion indicating magnitude and orientation of force vectors ($\alpha_m$=50°). Black lines indicate the instantaneous position of the wing at 40 temporally equidistant points during a stroke cycle. Small circles mark the leading edge. Time moves right to left during forth stroke, left to right during back stroke. Red vectors indicate instantaneous forces.

Now the vertical force is simply the lift and the drag only contribute to the horizontal force. As seen from Figure 17a, the wing could only produces a very small lift but a much larger drag; the force vector on the wing is approximately back-pointing in forth sweep and forward-pointing in the back sweep (Figure 17a). The mean lift, or mean vertical force, is only about 1/3 of that produced by using the real wing kinematics.

To explain why the lift is very small while drag very large, we plot the contributions to the lift and drag from the pressure and the viscous stress on the wing surface (Figure 18). It is seen that at this low *Re* (10.6), the viscous stress on the wing surface is rather large; and it make a large negative contribution to the lift and a large positive contribution to the drag, resulting in the very small lift and the much larger drag.

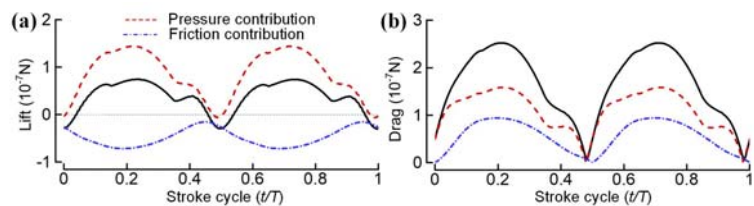

FIGURE 18. Contributions to the lift and drag from the pressure and the viscous stress on the wing surface ($\alpha_m$=45°).

The above results show that the tiny wasps cannot use wing kinematics commonly employed by the larger insects (wings sweeping back and forth in a horizontal plane) to generate the required vertical force. This explains why they use the special flapping motions, rowing and 'fling': these motions enable them to overcome the problem of insufficient lift at very small Reynolds number encountered by the commonly used flapping kinematics.

Finally, let us have some discussion on using drag principle for flight. The oar of a boat is in water during the effective stroke but in air during the return stroke, so that the return stroke requires little power. For the wasp, as seen in Figure 10a, no useful vertical force is produced in the period of 'clap' and moving upward and neither in the period of forth sweeping. These two periods may be taken as the "return stroke" if where the power requirement is very small. This is indeed the case: From data in Fig. 16, the power required during the rowing and the 'fling' periods are 60% and 21% of the total power requirement, respectively. However, in the two "return periods" ('clap' and moving upward; forth sweeping), the values are much smaller, only 10% and 9%, respectively. This is because in these two "return periods" the velocity and angle of attack of the wings are small (see Figure 11e); the angle of attack in the period of 'clap' and moving upward is almost zero (the wing surface is almost vertical). That small insects may use a drag mechanism for flight was suggested a long time ago (Horridge 1956) and also recently (Jones et al. 2015). But people had never yet observed a very small insect in flapping flight which used drag rather than lift. Here we for the first time observed very small insects using drag to produce vertical force for flight, and explained how the drag and the net vertical force were produced.

## 7. On the study limitations

The present study uses the method of computational fluid dynamics to assess the flows around and the aerodynamic forces acting on the insects. Two simplifying assumptions are made on the wing model used in the computation: (1) the fore- and hind-wings are modeled as a single wing. (2) the brim hairs that line the forewing (or hindwing) are modeled as an extension of the membranous wing. These are the limitations of the present study.

Some discussions on the validity of these simplifying assumptions are made here. A forewing or hind-wing of the *E. formosa* consists of a membranous part and the brim of marginal hairs (Figure 2a). In our model wing, the brim-hair part is assumed to be an extension of the membranous part. The membranous parts account for about 62% of the model-wing area and the brim-hair part accounts for about 38%. Sunada et al. (2002) measured the lift and drag forces on a bristled model wing and a solid-plate wing of the same shape. Different wing-motion patterns were considered: azimuthally rotating (revolving) at a constant angular speed and at a constant angular acceleration; translating at a constant speed and at a constant acceleration. Among these motions, the revolving motions at constant speed and at constant acceleration are the major components of the flapping motion of an insect wing. They showed that at $Re$ around 10, the aerodynamic forces acting on the bristled model wing were a little smaller than those on the solid wing. The ratio of aerodynamic forces between the bristled wing and the solid wing is between 0.85 and 1 for revolving at constant speed (see Fig. 4 in Sunada et. al. 2002) and between 0.7 and 1 for revolving at constant acceleration (see Fig. 8 in Sunada et. al. 2002). Let us assume that the ratio is 0.85, the mean value of 0.7 and 1. Using this value, the ratio of the aerodynamic forces between the real wing (membrane parts lined with brim marginal hairs) and the model wing (the brim hairs modeled as an extension of the membranous parts) can be estimated as: $0.85 \times 38\% + 62\% \approx 0.94\%$. That is, the simplifying assumption would cause an error of about 6% in aerodynamic force calculation. Furthermore, the distance between the bristles in the *E. formosa* wing (about $0.007R$, $R$ is the wing length) is much smaller than that in the bristled model wing in Sunada et al.'s experiment ($0.018R$); therefore, the error will be smaller than 6%. The gap between the forewing and the hindwing is full of the marginal hairs, and therefore, the above discussion also provide support to the assumption that the fore- and hind-wings can be modeled as a single wing.

The above discussions show that our model wing is a reasonably good approximation of the real wing. There is another point indicating that the model is a good approximation: the computed vertical force is approximately equal to the insect weight, a condition required for balanced flight.

## 8. Conclusion

The smallest insects are of very large number and of great ecological and biological significance. For these insects, the effect of air viscosity is so large that lift-generation mechanisms used by their larger counterparts become ineffective. How the weight-supporting force is generated is unknown. Our wing kinematics measurement and flow analysis in the tiny wasp *Encarsia formosa* in hovering flight answer the above question. The tiny insects perform two unusual wing-motions. One is "rowing": the wings move fast downward and backward, like stroking oars; the other is the previously discovered Weis-Fogh 'fling'. The rowing produces 70% of the required vertical force and the Weis-Fogh 'fling' the other 30%. The oaring wing mainly produces an approximately up-pointing drag, resulting in the vertical force. Because each oaring produces a starting flow, the drag is unsteady in nature and much greater than that in steady motion at the same velocities and angles of attack. Furthermore, our computation shows that if the tiny wasps employed the usual wing kinematics of the larger insects (flapping back and forth in a horizontal plane), vertical force produced would be only 1/3 of that by the real wing kinematics; i.e. they must use the special wing movements to overcome the problem of large viscous effects encountered by the commonly used flapping kinematics. We for the first time observe very small insects using drag to support their weight and explained how a net vertical force is generated when drag principle is applied.

**Acknowledgements**

This research was supported by grants from the National Natural Science Foundation of China (11672023, 11721202). X.C. was supported by the Academic Excellence Foundation of BUAA for PhD Students.

**Appendix A. Method of wing-kinematics measurement**

The method used to extract the 3D body and wing kinematics from the filmed data was a modification of that used in several previous works of our group (Liu & Sun 2008; Mou, Liu & Sun 2011). First the original method is outlined (see Mou, Liu & Sun 2011 for its detailed description). The body and wings were represented by models: the model of a wing was a flat-plate wing having the outline obtained by scanning the cut-off wing of the insect and the model of the body was two lines perpendicular to each other, which were the line connecting the head and the end of the abdomen and the line connecting the two wing hinges (Figure A1a). An interactive graphic user interface was developed using Matlab (Matlab v. 7.1, The Mathworks, Inc., Natick, MA, USA) to extract the 3D body and wing positions from the frames recorded by the three cameras. The positions and orientations of the models of the body and wings were adjusted until the best overlap between a model image and the displayed frame was achieved in three views, and at this point the positions and orientations of these models were taken as the positions and orientations of the body and the wings; the fitting process was manually done.

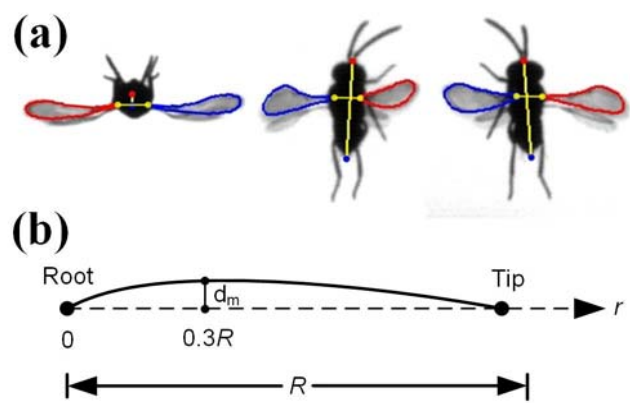

FIGURE A1. Kinematics extraction and wing bending definition. (a) Extraction of body and wing kinematics. (b) Definition of the spanwise bending of wing. $d_m$, maximum bending displacement, assumed to be at $0.3R$.

In the above method, the model of the wing was a rigid flat-plate. But for *E.*

*formosa*, considerable spanwise bending of the 'stalked' wing exists, as observed from the snapshots in Figure 4. The bending is relatively large during the rowing and the 'clap-and-fling' periods and smaller in the rest part of the cycle. We dealt with this problem by making the following changes on the above method. The model of the wing was no longer a rigid flat-plate wing, but a plate-wing that has spanwise bending (Figure A1b). Based on the observations, it was assumed that the maximum bending displacement, $d_m(t)$, was at 30% of wing length from the wing root. We approximate the bending displacement between $r/R=0$ and $r/R=0.3$ by a quadratic curve and that between $r/R=0.3$ and $r/R=1$ by another one, thus the bending displacement is determined when $d_m(t)$ determined. $d_m(t)$ is determined in process of matching the model image and the displayed frame. With the above modification, in addition to the three Euler angles, $d_m(t)$ was also measured.

**Appendix B. Measured wing kinematics and computed forces for all the eight individuals**

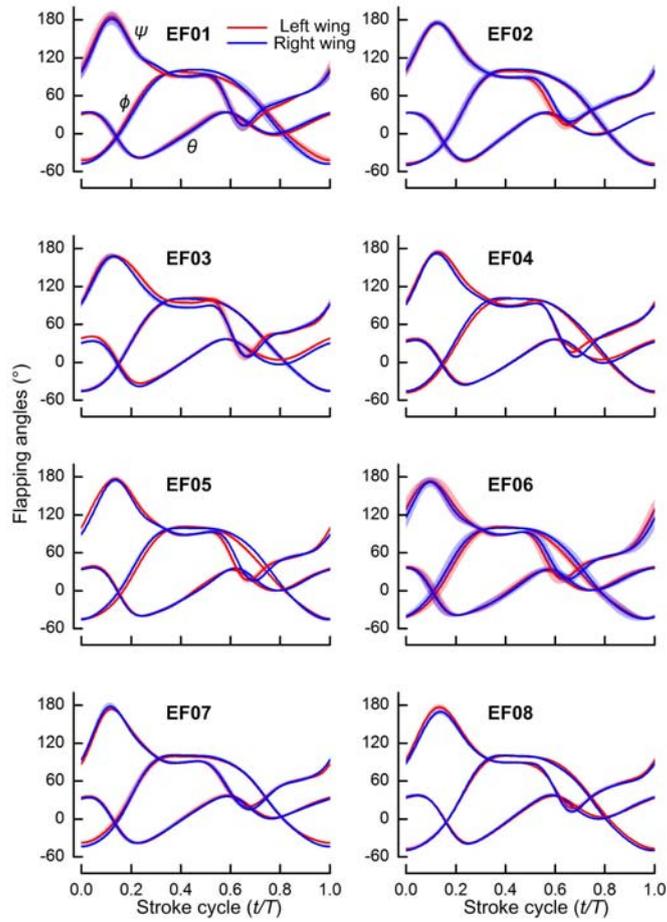

FIGURE A2. Wing kinematics for each of the eight wasps. Three flapping angles of wing (mean ± s.d.). For details, see the legend of Fig. 6a. Time courses of the angles among the eight individuals are similar.

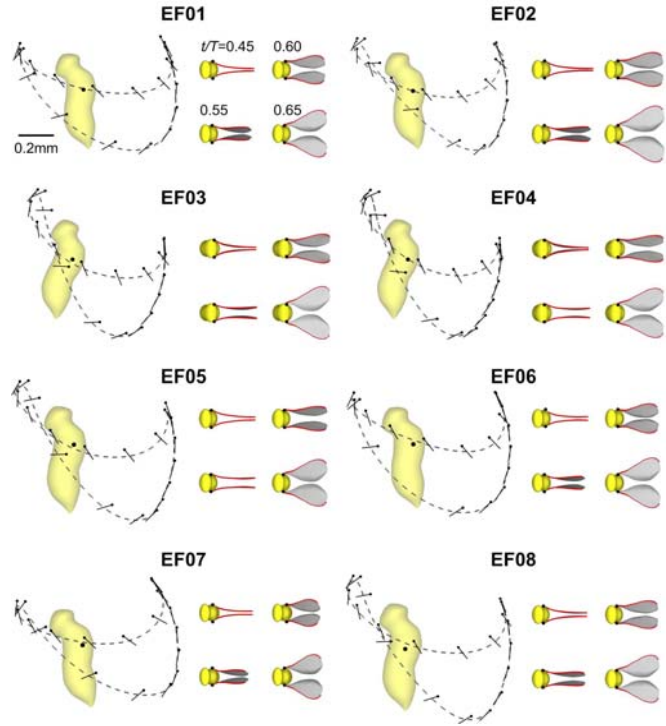

FIGURE A3. Stroke diagrams for each of the eight wasps. For each individual, left, wing-tip trajectory and orientation of wing at 20 temporally equidistant points; right, wing motion at dorsal stroke reversal, showing the 'clap-and-fling'. For details, see the legend of Fig. 7. Kinematic patterns for the eight individuals are approximately the same.

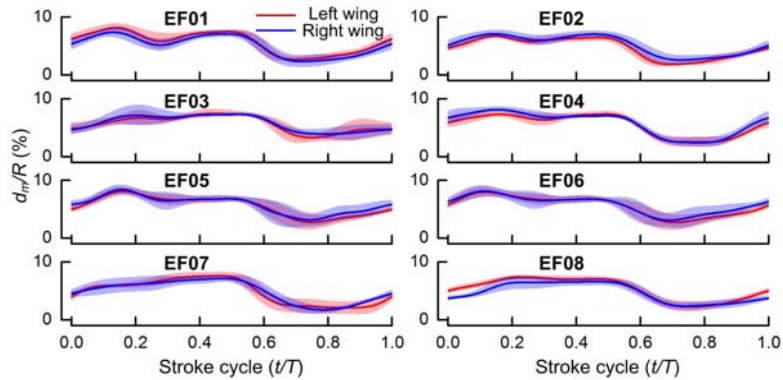

FIGURE A4. Spanwise bending of each of the eight wasps. The maximum bending displacement, $d_m/R$ (mean ± s.d.) in one stroke cycle. For details, see the legend of Fig. 6b.

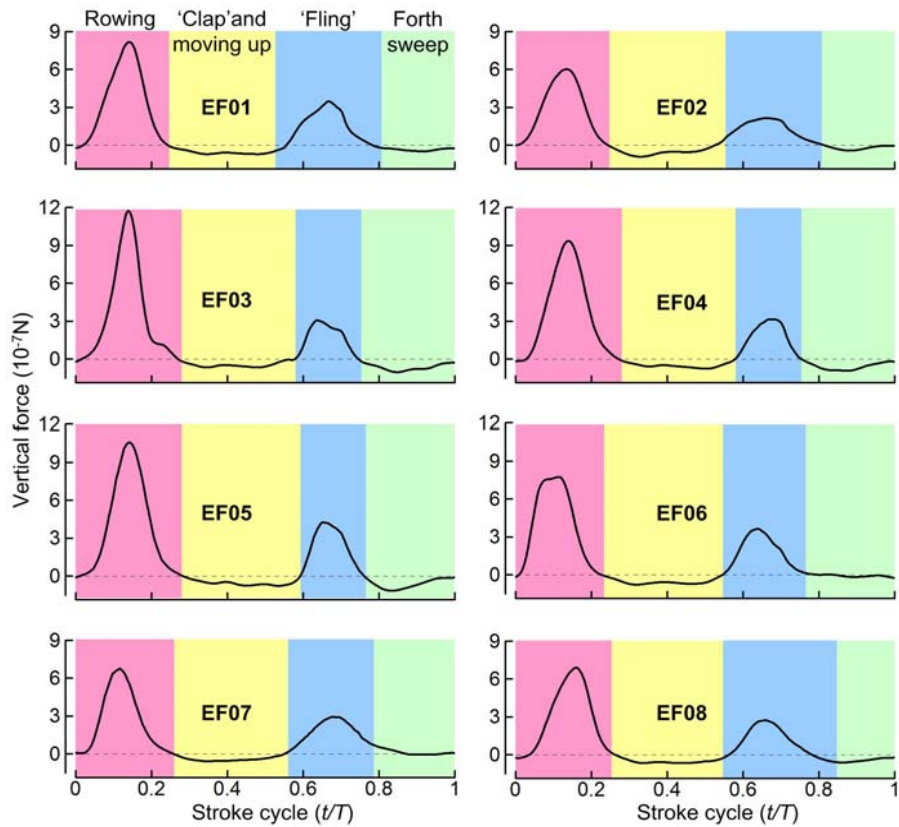

FIGURE A5. Vertical force for each of the eight wasps. The computed forces show great consistency for the eight individuals.

**REFERENCES**

ARORA, N., GUPTA, A., SANGHI, S., AONO, H. & SHYY, W. 2014 Lift-drag and flow structures associated with the 'clap and fling' motion. *Phys. Fluids* **26**, 071906.

BIRCH, J. M., DICKSON, W. B. & DICKSON, M. H. 2004 Force production and flow structure of the leading edge vortex on flapping wings at high and low Reynolds numbers. *J. Exp. Biol.* **207**, 1063-1072.

BOMPHREY, R. J., SRYGLEY, R. B., TAYLOR, G. K. & THOMAS, A. L. R. 2002 Visualizing the Flow around Insect Wings. *Phys. Fluids* **14**, S4.

CARR, Z. R., CHEN, C. & RINGUETTE, M. J. 2013 Finite-span rotating wings: three-dimensional vortex formation and variations with aspect ratio. *Exp. Fluids* **54**, 1-26.

DICKINSON, M. H., LEHMANN, F. O. & SANE, S. P. 1999 Wing rotation and the aerodynamic basis of insect flight. *Science* 284 (5422), 1954-1960.

DU, G. & SUN, M. 2012 Aerodynamic effects of corrugation and deformation in flapping wings of hovering hoverflies. *J. Theor. Biol.* **300**, 19-28.

DUDLEY, R. 2002 *The Biomechanics of Insect Flight: Form, Function, Evolution. Princeton University Press*.

ELLINGTON, C. P. 1984*a* The aerodynamics of hovering insect flight. I. The quasi-steady analysis. *Phil. Trans. R. Soc.* B **305**(1122), 1-15.

ELLINGTON, C. P. 1984*b* The aerodynamics of hovering insect flight. III.

Kinematics. *Phil. Trans. R. Soc*. B **305**(1122), 41-78.

ELLINGTON, C. P. 1984*c* The aerodynamics of hovering insect flight. IV. Aerodynamic mechanisms. *Phil. Trans. R. Soc*. B **305**(1122), 79-113.

ELLINGTON, C. P. 1984*d* The aerodynamics of hovering insect flight. VI. Lift and power requirements. *Phil. Trans. R. Soc*. B **305**(1122), 145-181.

ELLINGTON, C. P., VAN DEN BERG, C., WILLMOTT, A. P. ELLINGTON, C. P. & Thomas A L R. 1996 Leading-edge vortices in insect flight. *Nature*, **384**, 626-630.

GARMANN, D. J., VISBAL, M. R. & ORKWIS, P. D. 2013 Three-dimensional flow structure and aerodynamic loading on a revolving wing. *Phys. Fluids* **25**, 034101.

HAN, J.-S., CHANG, J. W. & HAN, J.-H. 2016 The advance ratio effect on the lift augmentations of an insect-like flapping wing in forward flight. *J. Fluid Mech.* **808**, 485-510.

HARBIG, R. R., SHERIDAN, J. & THOMPSON, M. C. 2013 Reynolds number and aspect ratio effects on the leading-edge vortex for rotating insect wing planforms. *J. Fluid Mech.* **717**, 166-192.

HILGENSTOCK, A. 1988 A fast method for the elliptic generation of three dimensional grids with full boundary control. *Num. Grid Generation in Computational Fluid Mechanics*, Swansea, UK, (ed. Sengupta, S., Hauster, J., Eiseman, P. R. & Thompson, J. F.) Pineridge Press, Ltd., 137-146.

HORRIDGE, G. A. 1956 The flight of very small insects. *Nature* **178,** 1334-1335.

HOWE, M. S. 1989 On unsteady surface forces, and sound produced by the normal chopping of a rectilinear vortex, *J. Fluid Mech.* **206**, 131-153.

HOWE, M. S. 1995 On the force and moment on a body in an incompressible fluid, with application to rigid bodies and bubbles at high and low Reynolds numbers, Quart. J. Mech. Appl. Math., **48**, 401-426.

JARDIN, T., FARCY, A. & DAVID, L. 2012 Three-dimensional effects in hovering flapping flight. *J. Fluid Mech.* **702**, 102-125.

JONES, S. K., LAURENZA, R., HEDRICK, T. L., GRIFFITH, B. E. & MILLER, L. A. 2015 Lift vs. drag based mechanisms for vertical force production in the smallest flying insects. *J. Theor. Biol.* **384**, 105-120.

KIM, D. & GHARIB, M. 2010 Experimental study of three-dimensional vortex structures in translating and rotating plates. *Exp. Fluids* **49**, 329-339.

LENTINK, D. & DICKINSON, M. H. 2009 Rotational accelerations stabilize leading edge vortices on revolving fly wings. *J. Expl Biol.* **212**, 2705-2719.

LIU, H., ELLINGTON, C. P., KAWACHI, K., VAN DEN BERG, C. & WILLMOTT, A. P. 1998 A computational fluid dynamic study of hawkmoth hovering. *J. Exp. Biol.* **201**, 461-477.

LIU L. G. & SUN, M. 2018 The added mass forces in insect flapping wings. *J. Theor. Biol.* **437**, 45–50.

LIU, Y. P. & SUN, M. 2008 Wing kinematics measurement and aerodynamics of hovering droneflies. *J. Exp. Biol.* **211**, 2014-2025.

LUO, G. Y. & SUN, M. 2005 The effects of corrugation and wing planform on the aerodynamic force production of sweeping model insect wings. *Acta Mech. Sin.* **12**, 1-7.

MANAR, F. & JONES, A. R. 2014 The Effect of Tip Clearance on Low Reynolds Number Rotating Wings. *52nd Aerosp. Sci. Meet.* 1452.

MENG, X. G. & SUN, M. 2013 Aerodynamic effects of wing corrugation at gliding flight at low Reynolds numbers. *Phys. Fluids* 25(7), 071905.
MENG, X. G., & SUN, M. 2015 Aerodynamics and vortical structures in hovering fruitflies. *Phys. Fluids* **27**, 031901.
MILLER, L. A. & PESKIN, C. S. 2004 When vortices stick: An aerodynamic transition in tiny insect flight. *J. Exp. Biol.* **207**, 3073-3088.
MILLER, L. A. & PESKIN, C. S. 2005 A computational fluid dynamics of 'clap and fling' in the smallest insects. *J. Exp. Biol.* **208**, 195-212.
MENG, X. G. & SUN, M. 2013 Aerodynamic effects of wing corrugation at gliding flight at low Reynolds numbers. *Phys. Fluids* **25**, 071905.
MOU, X. L., LIU, Y. P. & SUN, M. 2011 Wing motion measurement and aerodynamics of hovering true hoverflies. *J. Expl Biol.* **214**, 2832-2844.
OZEN, C. A. & ROCKWELL, D. 2012 Three-dimensional vortex structure on a rotating wing. *J. Fluid Mech.* **707**, 541-550.
ROGERS, S. E., KWAK, D. & KIRIS, C. 1991 Steady and unsteady solutions of the incompressible Navier-Stokes equations. *AIAA J.* **29**, 603-610.
ROGERS, S. E., & PULLIAM, T. H. 1994 Accuracy enhancements for overset grids using a defect correction approach. *AIAA Paper.* 94-0523.
SANE, S. P. 2003 The aerodynamics of insect flight. *J. Exp. Biol.* **206**, 4191-4208.
SHYY, W., AONO, H., CHIMAKURTHI, S. K., TRIZILA, P., KANG, C. K., CESINK, C. E. S. & LIU, H. 2010 Recent progress in flapping wing aerodynamics and aeroelasticity. *Prog. Aerosp. Sci.* **46**, 284-327.
SHYY, W., TRIZILA, P., KANG, C. K. & AONO, H. 2009 Can tip vortices enhance lift of a flapping wing? *AIAA J.* **47**, 289-293.
SPEDDING, G. R. & MAXWORTHY, T. 1986 The generation of circulation and lift in a rigid two-dimensional fling. *J. Fluid Mech.* **165**, 247-272.
SUN, M. & DU, G. 2003 Lift and power requirements of hovering insect flight. *Acta Mech. Sin.* **19**, 458-469.
SUN, M. & LAN, S. L. 2004 A computational study of the aerodynamic forces and power requirements of dragonfly (Aeschna juncea) hovering. *J. Exp. Biol.* **207**, 1887-1901.
SUN, M. & YU, X. 2003 Flows Around Two Airfoils Performing Fling and Subsequent Translation and Subsequent Clap. *Acta Mech. Sin.* **19**, 103–117.
SUN, M. & YU, X. 2006 Aerodynamic Force Generation in Hovering Flight in a Tiny Insect. *AIAA J.* **44,** 1532-1540.
SUN, M. & WU, J. H. 2004 Large aerodynamic forces on a sweeping wing at low Reynolds number. *Acta Mech. Sin.* **20**, 24-31.
SUNADA, S., TAKASHIMA, H., HATTORI, T., YASUDA, K. & KAWACHI, K. 2002 Fluid-dynamic characteristics of a bristled wing. *J. Exp. Biol.* **205,** 2737-2744.
WAKELING, J. M. & ELLINGTON, C. P. 1997 Dragonfly flight. II. Velocities, accelerations and kinematics of flapping flight. *J. Exp. Biol.* **200**, 557-582.
WANG, Z. J. 2004 The role of drag in insect hovering. *J. Exp. Biol.* **207**, 4147-4155.
WANG, Z. J. 2005 Dissecting insect flight. *Annu. Rev. Fluid Mech.* **37**, 183-210.
WEIS-FOGH, T. 1973 Quick estimates of flight fitness in hovering animals, including novel mechanisms for lift production. *J. Exp. Biol.* **59**, 169-203.
WEIS-FOGH, T. 1975 Flapping flight and power in birds and insects, conventional and novel mechanisms. In Swimming and Flying in Nature (ed. Wu, T. Y., Brokaw, C. J. & Brennen, C.), vol. 2, pp. 729-762. Plenum, New York.
WHITE, F. M. 1991 Viscous fluid flow, Vol. 2, Boston: McGraw-Hill Higher Education.

WOJCIK, C. J. & BUCHHOLZ, J. H. J. 2014 Vorticity transport in the leading-edge vortex on a rotating blade. *J. Fluid Mech.* **743**, 249-261.

WOLFINGER, M. & ROCKWELL, D. 2014 Flow structure on a rotating wing: effect of radius of gyration. *J. Fluid Mech.* **755**, 83-110.

WU, J. H. & SUN, M. 2004 Unsteady aerodynamic forces of a flapping wing. *J. Exp. Biol.* **207,** 1137-1150.

ZHU, H. J. & SUN, M. 2017 Unsteady aerodynamic force mechanisms of a hoverfly hovering with a short stroke-amplitude. *Phys. Fluids* **29**, 081901.